\documentclass{l4dc2027}

\theorembodyfont{\normalfont}

\newtheorem{proposition}{Proposition}
\newtheorem{definition}{Definition}

\newtheorem{corollary}{Corollary}
\newtheorem{lemma}{Lemma}
\newtheorem{theorem}{Theorem}
\newtheorem{remark}{Remark}

\title[Parametric Data-Driven Hamiltonian]{Data-Driven Hamiltonian Under Parametric Uncertainty}
\usepackage{times}
\usepackage{booktabs}

\author{%
 \Name{Zach Liu} \Email{zachliu@ucla.edu}\\
 \addr University of California, Los Angeles
 \AND
 \Name{Namhoon Cho} \Email{namhoon.cho@snu.ac.kr}\\
 \addr Seoul National University
 \AND
 \Name{Jason J. Choi} \Email{jjhchoi@ucla.edu}\\
 \addr University of California, Los Angeles
}

\begin{document}

\maketitle

\begin{abstract}%
Hamilton--Jacobi (HJ) reachability computes the states from which a system can be kept out of an
unsafe region, but it requires knowledge of dynamics. The data-driven Hamiltonian (DDH), under a black-box dynamics setting, lower bounds the
Hamiltonian directly from trajectory data, which produces a guaranteed conservative safe set. However, it needs a Lipschitz constant of the dynamics, and its bound loosens with the distance from the data. We consider dynamics that are linear in unknown parameters through a known feature vector, with a Gaussian prior on the parameters, and use this
structure to construct a parametric DDH. From a few short trajectories with noisy velocity measurements, the parametric DDH lower-bounds the
Hamiltonian at all states and costates with a probability of at least $1-\delta$. The safe set obtained by replacing the Hamiltonian with this bound is then contained in the true safe set with the same probability.
For control-affine systems, evaluating the bound amounts to a second-order cone program, so it can be solved efficiently within existing grid-based and neural PDE solvers. We demonstrate the approach on a two-state system and a six-state lunar lander.%
\end{abstract}


\section{Introduction}

Autonomous systems such as drones, aircraft, and self-driving vehicles operate in environments
shared with people, where a single unsafe action can cause injuries. Safety analysis determines which initial states allow the system to remain outside an unsafe region. Hamilton--Jacobi (HJ) reachability computes this safe set and an associated safety controller for general nonlinear systems \citep{mitchell2005time,bansal2017hamilton}, and neural PDE solvers extend the computation to higher-dimensional systems \citep{bansal2021deepreach}. In these frameworks, the dynamics enter the reachability computation only through the Hamiltonian, and in practice, they are known only approximately. A safe set computed from an inaccurate model can contain states that are in fact unsafe.

Robust reachability treats model error as a bounded disturbance \citep{mitchell2005time}, and learning-based methods estimate this disturbance from data with Gaussian process regression \citep{akametalu2014reachability,fisac2019general}. \citet{choi2025ddh} instead introduce the data-driven Hamiltonian (DDH), a lower bound on the Hamiltonian constructed directly from trajectory data, which yields an inner approximation of the safe set without identifying or learning a model. This original DDH formulation requires a Lipschitz constant of the dynamics, and its uncertainty penalty grows with the distance from each data point, making the lower bound more conservative. Thus, a large Lipschitz constant or sparse data leads to an overly conservative safe set.


Modeling the uncertainty as linear in unknown parameters gives a useful structure for the analysis and control of uncertain systems, and many adaptive control methods are developed on this representation \citep{slotine1987adaptive}, including adaptive control barrier functions for safety \citep{taylor2020adaptive,lopez2021robust}. In many systems, the structure of the model equations follows from first principles or physics laws, and only physical parameters, such as inertial and thrust parameters, are uncertain. Rigid-body equations of motion, for example, are linear in the inertial parameters \citep{atkeson1986estimation}. Otherwise, the features can be chosen from a library of candidate functions \citep{brunton2016discovering} or meta-learned from related tasks \citep{richards2021adaptive,oconnell2022neural}, possibly with a prior over the parameters \citep{harrison2018meta}.

Our goal is to use this linear-in-parameter structure to bound and approximate the
Hamiltonian from data without access to the Lipschitz constant of the dynamics. We call this bound the \textit{parametric data-driven Hamiltonian}. It replaces the Lipschitz constant of the DDH with known features and a Gaussian prior. The Hamiltonian maximizes an objective that is linear in the unknown parameters. Bayesian linear regression infers the parameter posterior from short trajectories with noisy velocity measurements. Maximizing the posterior mean of this objective minus a multiple of its posterior standard deviation yields the parametric DDH, a lower bound on the Hamiltonian that holds with probability at least $1-\delta$. Since the features are known functions of the state and control, the parameter posterior inferred from the data determines the bound at every state, including states far from the samples. 

On a two-state system, the computed safe sets lie inside the true safe set, whereas omitting the posterior uncertainty term can label unsafe states as safe. For a six-state lunar lander, Monte Carlo evaluation provides empirical evidence that the data-driven safe set computed using the neural reachability solver \citep{bansal2021deepreach} with the parametric DDH conservatively approximates the true safe set.

This paper makes three contributions. First, we propose the parametric DDH for dynamics that are linear in unknown parameters, which lower-bounds the true Hamiltonian at all states and costates with probability at least $1-\delta$, without a Lipschitz constant. Second, for control-affine systems, we formulate its evaluation as a second-order cone program, so that it can be efficiently solved within existing HJ solvers. Third, on a two-state system and a six-state lunar lander, we show that a few short trajectories suffice to compute safe sets that are validated against the true dynamics.

\section{Problem Formulation and Background}\label{sec:problem}

\subsection{Problem Formulation}
We consider the continuous-time, nonlinear system
\begin{equation}\label{eq:system}
  \dot{\boldsymbol{x}}(s)=f\big(\boldsymbol{x}(s),\boldsymbol{u}(s)\big)\quad\text{for } s\in[-t,0],
  \qquad \boldsymbol{x}(-t)=x,
\end{equation} where we set the initial time as $-t$ and the initial state as $x$, with state $\boldsymbol{x}(s)\in\mathbb{R}^n$, and control $\boldsymbol{u}(s)\in U\subset\mathbb{R}^m$. $U$ is assumed to be compact and convex. The dynamics are linear in an unknown parameter
matrix $\Theta\in\mathbb{R}^{d\times n}$,
\begin{equation}\label{eq:greybox}
  f(x,u)=\Theta^\top\phi(x,u),
\end{equation}
where the feature vector $\phi:\mathbb{R}^n\times U\to\mathbb{R}^d$ is known and globally Lipschitz continuous. For mechanical systems, $\phi$ can often be derived from first principles, since the Euler--Lagrange equations of motion are linear in physical parameters such as masses, inertias, and 
thrusts~\citep{slotine1991applied}, and the state equation remains linear in these parameters or their ratios when the inertia matrix is known or constant and diagonal. Otherwise, $\phi$ can be meta-learned from data of related systems that share $\phi$ but differ in $\Theta$~\citep{harrison2018meta, richards2021adaptive, oconnell2022neural}.
We additionally denote
the vectorization of $\Theta$ as $\tilde\Theta:=\mathrm{vec}(\Theta)\in\mathbb{R}^{dn}$.

Two sources of information about $\Theta$ are available: (i) a Gaussian prior
$\tilde\Theta\sim\mathcal{N}(m_0,\Sigma_0)$, and (ii) a trajectory dataset
$\mathcal{D}=\{(x^i,u^i,v^i)\}_{i=1}^N$,
where $v^i$ is a noisy measurement of the state velocity,
\begin{equation}\label{eq:measurement}
  v^i=f(x^i,u^i)+\varepsilon^i,\qquad \varepsilon^i\sim\mathcal{N}(0,\sigma^2 I_n).
\end{equation}
The noise terms $\varepsilon^i$ are independent and identically distributed.

Given the prior and $\mathcal{D}$, we wish to calculate which states can keep the system away from an unsafe
set $\mathcal{X}_{\mathcal{U}}\subset\mathbb{R}^n$ over a horizon $t\ge 0$. These states form the
avoid backward reachable tube (BRT), which we call the safe set:
\begin{equation}\label{eq:safe-set}
  \mathcal{S}(t):=\big\{x\in\mathbb{R}^n\mid\exists\,\boldsymbol{u}(\cdot)\ \text{s.t.}\
  \forall s\in[-t,0],\ \boldsymbol{x}(s)\notin\mathcal{X}_{\mathcal{U}}\big\}.
\end{equation}
From any state outside $\mathcal{S}(t)$, the trajectory reaches $\mathcal{X}_{\mathcal{U}}$ within
the horizon regardless of the control signal and becomes unsafe.

\subsection{Hamilton-Jacobi Reachability}
We analyze the safety of system \eqref{eq:system} with Hamilton--Jacobi (HJ) reachability
\citep{bansal2017hamilton}, which obtains the safe set from the solution of a partial
differential equation. We write $\mathcal{X}_{\mathcal{U}}=\{x\mid\ell(x)< 0\}$, where the
constraint function $\ell$ is bounded and Lipschitz continuous.

HJ reachability characterizes $\mathcal{S}(t)$
through the sign of the value function
\begin{equation}\label{eq:value}
  V(x,t):=\sup_{\boldsymbol{u}(\cdot)}\ \min_{s\in[-t,0]}\ \ell\big(\boldsymbol{x}(s)\big)
  \quad\Rightarrow\quad \mathcal{S}(t)=\{x\mid V(x,t)\ge0\}.
\end{equation}
By the dynamic programming principle, for any $T>0$, $V$ is the unique bounded continuous viscosity
solution on $\mathbb{R}^n\times[0,T]$ of the Hamilton-Jacobi variational inequality (HJ-VI)
\begin{equation}\label{eq:hjvi}
  0=\min\Big\{\ell(x)-V(x,t),\ -\partial_t V(x,t)+H\big(x,\nabla_x V(x,t)\big)\Big\},
\end{equation}
with initial condition $V(x,0)=\ell(x)$
\citep{fisac2015reach,altarovici2013general}. 
Note that $t\ge0$ denotes the horizon length and the corresponding trajectories evolve over $s\in[-t,0]$. The Hamiltonian in \eqref{eq:hjvi} is defined as:
\begin{equation}\label{eq:hamiltonian}
  H(x,p):=\max_{u\in U}\ p^\top f(x,u).
\end{equation}

Evaluating \eqref{eq:hamiltonian} requires knowledge of $f$. Our goal is to construct an inner
approximation of the safe set from the prior and a small dataset. Specifically, given
$\delta\in(0,1)$ and $T>0$, we seek to verify a set in time, $\hat{\mathcal{S}}(t)$, such that
\begin{equation}\label{eq:goal}
  \mathbb{P}\big(\hat{\mathcal{S}}(t)\subseteq\mathcal{S}(t),\ \forall t\in[0,T]
  \ \big|\ \mathcal{D}\big)\ \ge\ 1-\delta,
\end{equation}
where the probability is with respect to the posterior distribution of $\Theta$.

\section{Parametric Data-Driven Hamiltonian}\label{sec:method}

In this section, we construct the parametric DDH, a lower bound on the Hamiltonian that holds with probability at least $1-\delta$. We first use the linear-in-parameter structure to infer a Gaussian posterior of the parameters from the data (Section~\ref{sec:posterior}). We then use a confidence region of this posterior to bound the Hamiltonian from below at all states and costates simultaneously, which yields a lower approximation of the value function and thus, an inner approximation of the safe set (Section~\ref{sec:bound}). Finally, we show that, for control-affine systems, the bound can be evaluated efficiently as a second-order cone program (Section~\ref{sec:socp}).

\subsection{Linear-Gaussian Model of the Data}\label{sec:posterior}

For any costate $p\in\mathbb{R}^n$, the identity $\mathrm{vec}(\phi p^\top)=p\otimes\phi$ gives
\begin{equation}\label{eq:integrand}
\begin{aligned}
  h(x,u,p)&:=p^\top f(x,u)=\mathrm{tr}\big(\Theta^\top\phi(x,u)\,p^\top\big)
  =\tilde\Theta^\top\tilde\phi(x,u,p),\\
  \tilde\phi(x,u,p)&:=p\otimes\phi(x,u)\in\mathbb{R}^{dn},
\end{aligned}
\end{equation}
so that $H(x,p)=\max_{u\in U}h(x,u,p)$. $h(x,u,p)$ is linear in $\tilde\Theta$, and we will use this linearity to construct our probabilistic lower bound of the Hamiltonian.

The measurements are also linear in $\tilde\Theta$. With $\phi_i:=\phi(x^i,u^i)$ and
$A_i:=I_n\otimes\phi_i^\top\in\mathbb{R}^{n\times dn}$, we have $A_i\tilde\Theta=\Theta^\top\phi_i$,
so \eqref{eq:measurement} becomes $v^i=A_i\tilde\Theta+\varepsilon^i$. Stacking the $N$ samples
gives the linear-Gaussian model
\begin{equation}\label{eq:regression}
  y=A\tilde\Theta+\varepsilon,\qquad
  y:=\begin{bmatrix}v^1\\ \vdots\\ v^N\end{bmatrix},\quad
  A:=\begin{bmatrix}A_1\\ \vdots\\ A_N\end{bmatrix},\quad
  \varepsilon\sim\mathcal{N}(0,\Sigma_y),\quad \Sigma_y:=\sigma^2 I_{nN},
\end{equation}
and the posterior is Gaussian, $\tilde\Theta\mid\mathcal{D}\sim\mathcal{N}(m_D,\Sigma_D)$, with
\citep[Sec.~3.3]{bishop2006pattern}
\begin{equation}\label{eq:posterior}
  \Sigma_D=\big(\Sigma_0^{-1}+A^\top\Sigma_y^{-1}A\big)^{-1},\qquad
  m_D=\Sigma_D\big(\Sigma_0^{-1}m_0+A^\top\Sigma_y^{-1}y\big).
\end{equation}
Because $\Sigma_0\succ 0$ and $A^\top\Sigma_y^{-1}A\succeq 0$, we have $\Sigma_D\succ 0$. The
update does not require the explicit construction of $A$, since
$A^\top\Sigma_y^{-1}A=\sigma^{-2}I_n\otimes\sum_{i}\phi_i\phi_i^\top$ and
$A^\top\Sigma_y^{-1}y=\sigma^{-2}\,\mathrm{vec}\big(\sum_{i}\phi_i(v^i)^\top\big)$; its computation is
dominated by inverting the $dn\times dn$ matrix $(\Sigma_0^{-1}+A^\top\Sigma_y^{-1}A)$ in \eqref{eq:posterior}.

\subsection{A Probabilistic Lower Bound on the Hamiltonian}\label{sec:bound}

For $\delta\in(0,1)$, let $\chi^2_{dn,1-\delta}$ denote the $(1-\delta)$-quantile of the
chi-square distribution with $dn$ degrees of freedom, and define the radius and the confidence
ellipsoid
\begin{equation}\label{eq:ellipsoid}
  \beta_D:=\sqrt{\chi^2_{dn,1-\delta}},\qquad
  \mathcal{E}_D:=\big\{\tilde\Theta:\ \|\Sigma_D^{-1/2}(\tilde\Theta-m_D)\|_2\le\beta_D\big\}.
\end{equation}

\begin{lemma}\label{lem:ellipsoid}
$\mathbb{P}(\tilde\Theta\in\mathcal{E}_D\mid\mathcal{D})=1-\delta$, and if \;$\tilde\Theta\in\mathcal{E}_D$,
then for every $a\in\mathbb{R}^{dn}$,
\begin{equation}\label{eq:support}
  \tilde\Theta^\top a\ge m_D^\top a-\beta_D\,\|\Sigma_D^{1/2}a\|_2.
\end{equation}
\end{lemma}
\begin{proof}
Given $\mathcal{D}$, the vector $\xi:=\Sigma_D^{-1/2}(\tilde\Theta-m_D)$ is standard normal in
$\mathbb{R}^{dn}$, so $\|\xi\|_2^2$ is chi-square distributed with $dn$ degrees of freedom. By the
definition of $\beta_D$, $\mathbb{P}(\|\xi\|_2\le\beta_D\mid\mathcal{D})=1-\delta$, which is the first
claim. For the second, write $\tilde\Theta=m_D+\Sigma_D^{1/2}\xi$. For every $a\in\mathbb{R}^{dn}$, the
Cauchy--Schwarz inequality gives
\[
  \tilde\Theta^\top a=m_D^\top a+\xi^\top\Sigma_D^{1/2}a\ \ge\ m_D^\top a-\|\xi\|_2\,\|\Sigma_D^{1/2}a\|_2 .
\]
If $\tilde\Theta\in\mathcal{E}_D$, then $\|\xi\|_2\le\beta_D$ by \eqref{eq:ellipsoid}, and
\eqref{eq:support} follows.
\end{proof}

Applying \eqref{eq:support} to the objective function \eqref{eq:integrand} leads to the
\emph{parametric data-driven Hamiltonian}
\begin{equation}\label{eq:pddh}
  \hat H(x,p):=\max_{u\in U}\ \hat h(x,u,p),\qquad
  \hat h(x,u,p):=m_D^\top\tilde\phi(x,u,p)-\beta_D\,\big\|\Sigma_D^{1/2}\tilde\phi(x,u,p)\big\|_2.
\end{equation}

\begin{theorem}[Hamiltonian lower bound]\label{thm:hamiltonian}
In the setting of Section~\ref{sec:problem},
\begin{equation}\label{eq:thm-hamiltonian}
  \mathbb{P}\big(\hat H(x,p)\le H(x,p)\ \forall x\in\mathbb{R}^n,\ p\in\mathbb{R}^n
  \ \big|\ \mathcal{D}\big)\ \ge\ 1-\delta.
\end{equation}
\end{theorem}
\begin{proof}
On the event $\tilde\Theta\in\mathcal{E}_D$, Lemma~\ref{lem:ellipsoid} with $a=\tilde\phi(x,u,p)$ and
\eqref{eq:integrand} give $h(x,u,p)=\tilde\Theta^\top\tilde\phi(x,u,p)\ge\hat h(x,u,p)$ for all
$(x,u,p)$. Maximizing both sides over $u\in U$ gives $\hat H(x,p)\le H(x,p)$ for all $(x,p)$.
By Lemma~\ref{lem:ellipsoid}, this event has probability $1-\delta$.
\end{proof}

Building on this result, we show that the parametric DDH yields a conservative approximation of the safe set with high probability:

\begin{theorem}[Lower approximation of the value function]\label{thm:value}
Let $\hat V$ denote the continuous viscosity solution of \eqref{eq:hjvi} with $H$ replaced by
$\hat H$, and let $\hat{\mathcal{S}}(t):=\{x\mid\hat V(x,t)\ge 0\}$. With probability at least
$1-\delta$, $\hat V\le V$ on $\mathbb{R}^n\times[0,T]$ and $\hat{\mathcal{S}}(t)\subseteq\mathcal{S}(t)$
for all $t\in[0,T]$. Hence $\hat{\mathcal{S}}$ satisfies \eqref{eq:goal}.
\end{theorem}
\begin{proof}
On the event $\tilde\Theta\in\mathcal{E}_D$, we have $\hat H (x, p)\le H(x, p)$ for all $x, p \in \mathbb{R}^n$, as shown in the proof of
Theorem~\ref{thm:hamiltonian}, and the comparison principle in Appendix~\ref{app:hj-comparison}, Lemma~\ref{prop:hamiltonian-monotonicity} gives $\hat V(x, t)\le V(x,t)$ for all $(x,t)\in\mathbb{R}^n\times[0,T]$. If $\hat V(x,t)\ge0$, then
$V(x,t)\ge\hat V(x,t)\ge0$, so $\hat{\mathcal{S}}(t)\subseteq\mathcal{S}(t)$ for all $t\in[0,T]$.
By Lemma~\ref{lem:ellipsoid}, this event has probability $1-\delta$.
\end{proof}

\subsection{Evaluating \texorpdfstring{$\hat H$}{the Hamiltonian bound} as a Second-Order Cone Program for Control-Affine Systems}\label{sec:socp}

We now assume that the features are control-affine, $\phi(x,u)=a(x)+B(x)u$ with
$a:\mathbb{R}^n\to\mathbb{R}^d$ and $B:\mathbb{R}^n\to\mathbb{R}^{d\times m}$. Then $\tilde\phi$ is
also control-affine,
\begin{equation}\label{eq:affine}
  \tilde\phi(x,u,p)=\Phi(x,p)\,u+\tilde\phi_0(x,p),\qquad
  \Phi(x,p):=p\otimes B(x)\in\mathbb{R}^{dn\times m},\quad
  \tilde\phi_0(x,p):=p\otimes a(x).
\end{equation}
Let $R_D$ be any matrix with $R_D^\top R_D=\Sigma_D$, for example the transposed Cholesky
factor of $\Sigma_D$.

\begin{theorem}[SOCP form of $\hat H$]\label{thm:socp}
For every $(x,p)$, with $c:=\Phi^\top m_D$, $\alpha:=\tilde\phi_0^\top m_D$, $M:=R_D\Phi$ and
$r:=R_D\tilde\phi_0$,
\begin{equation}\label{eq:socp}
  \hat H(x,p)=\max_{u\in U,\ s\in\mathbb{R}}\ c^\top u+\alpha-\beta_D\,s
  \quad\text{subject to}\quad \|Mu+r\|_2\le s.
\end{equation}
This problem is a second-order cone program (SOCP) when $U$ is a second-order-cone representable convex set~\citep{lobo1998applications}.
\end{theorem}
\begin{proof}
By \eqref{eq:affine}, $m_D^\top\tilde\phi(x,u,p)=c^\top u+\alpha$. Since $R_D^\top R_D=\Sigma_D$,
we also have $\|\Sigma_D^{1/2}\tilde\phi(x,u,p)\|_2=\|R_D\tilde\phi(x,u,p)\|_2=\|Mu+r\|_2$. Hence
$\hat h(x,u,p)=c^\top u+\alpha-\beta_D\|Mu+r\|_2$. For a fixed $u\in U$, the constraint requires $s\ge\|Mu+r\|_2$. Since $\beta_D>0$, the objective
decreases as $s$ increases, so the best choice is the smallest feasible value, $s=\|Mu+r\|_2$.
Substituting it into the objective gives $c^\top u+\alpha-\beta_D\|Mu+r\|_2=\hat h(x,u,p)$. Taking
the maximum over $u\in U$ then gives $\hat H(x,p)$ by \eqref{eq:pddh}. The objective is linear in $(u,s)$, the constraint is a
second-order cone constraint on the affine map $(u,s)\mapsto(Mu+r,s)$, which
gives the last claim.
\end{proof}

\begin{figure}[!t]
  \centering
  \includegraphics[width=0.9\linewidth]{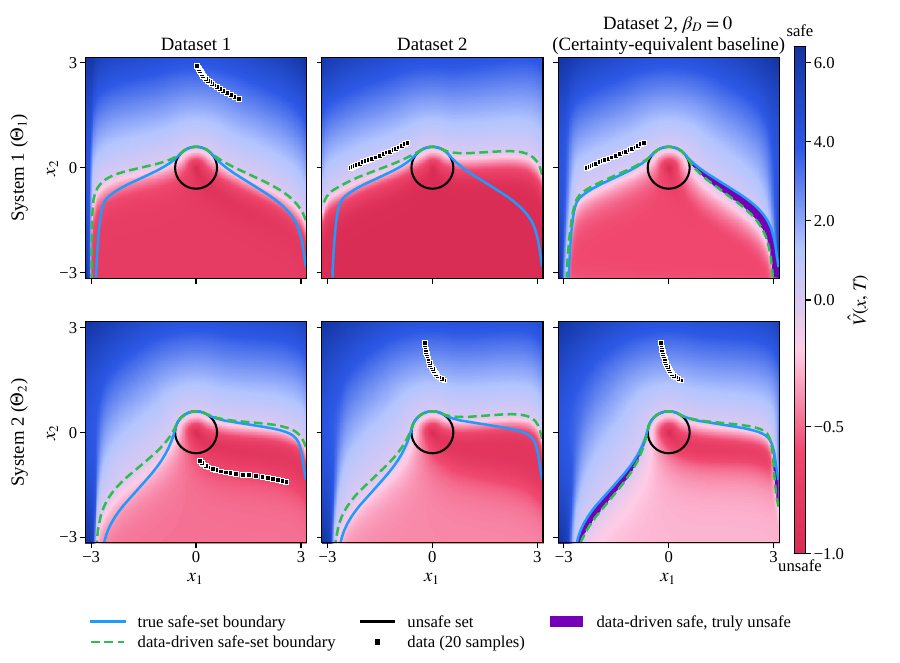}
  \vspace{-1.5em}
  \caption{Safe sets of the two-state system at $T=6$~s for two random parameter matrices (rows). First and
  second columns: the parametric data-driven Hamiltonian with Datasets~1 and~2; the data-driven safe sets (dashed) lie inside the true safe sets (solid). Third column: a certainty-equivalent baseline ($\beta_D=0$) that treats the posterior mean of $\Theta$ as exact; its safe set includes truly unsafe states (purple). The heat map shows $\hat V$.}

  \label{fig:toy}
\vspace{-1em}
\end{figure}

\begin{remark}
Under the assumed Bayesian model, the posterior remains well-defined since $\Sigma_0\succ0$, without requiring the persistency of excitation or a full-column-rank data matrix ($A$) condition. Insufficient excitation instead leaves greater posterior uncertainty in directions poorly informed by the data. This uncertainty enters the parametric DDH through the penalty ($\beta_D\|\Sigma_D^{1/2}\widetilde{\phi}(x,u,p)
\|_2$), increasing conservatism where those directions affect safety. Thus, the quality of data determines how conservative the resulting safety certificate is, and datasets of equal size can yield a different size and shape of the approximated safe sets, as we will see in our numerical experiment (Figure~\ref{fig:toy}).
\end{remark}

\section{Numerical Experiments}\label{sec:experiments}

We apply the parametric data-driven Hamiltonian to a two-state system and to a six-state lunar lander, with $\delta=0.1$ in both.
We compare each data-driven safe set with the safe set
computed by the same solver with the true Hamiltonian. For each dataset, initial states and controls are drawn uniformly from the state and control bounds.


For the two-state system, we solve \eqref{eq:socp} at all grid points in one batch on the GPU with the
conic solver Moreau \citep{barratt2026moreau} through cvxpylayers \citep{agrawal2019differentiable}. For the lunar lander, which
requires solving \eqref{eq:socp} at every collocation point in every training iteration, we run the
Adam optimizer \citep{kingma2015adam} for a fixed number of iterations and project each iterate onto
$U$, so the computed value does not exceed $\hat H(x,p)$ up to rounding error and the inequality of
Theorem~\ref{thm:hamiltonian} also holds for it.

\subsection{Two-State Example}\label{sec:toy}

The first system is
\begin{equation}\label{eq:toy}
  \dot x=\Theta^\top\phi(x,u),\qquad
  \phi(x,u)=\big[1,\ \sin x_1,\ (1+0.5\cos x_2)\,u\big]^\top,\qquad u\in[-1,1],
\end{equation}
with all six entries of $\Theta$ unknown, randomly selected from its prior distribution, and lying inside the confidence ellipsoid. The unsafe set is the disk of radius $0.6$ centered at the origin, with $\ell(x)=\min\{\|x\|_2/0.6-1,\,6.5\}$. We use the prior $\mathcal{N}(0,I_6)$, the noise level
$\sigma=0.1$, and datasets of one trajectory with 20 samples, yielding $\beta_D=3.26$. We compute
the value functions up to $T=6$~s on a $151\times 151$ grid over $[-\pi,\pi]^2$ with the level set
method \citep{mitchell2005time}.


Figure~\ref{fig:toy} shows the safe sets at $t=6$~s. In the first and second columns, each under a distinct dataset consisting of a single randomly generated trajectory, the data-driven safe
set is contained entirely inside the true safe set. The gap between the two boundaries contains states
that are safe but excluded from the data-driven safe set. In the third column, $\hat H$ uses only the
posterior mean to approximate the Hamiltonian, and the containment $\hat{\mathcal{S}}(t)\subseteq\mathcal{S}(t)$ fails: the purple regions capture unsafe states that are falsely labeled as safe.

\subsection{Lunar Lander}\label{sec:lander}

The second system is a six-state lunar lander with the state $x=(p_x,v_x,p_y,v_y,\theta,\omega)$ and the thruster inputs
$u=(u_0,u_1,u_2)\in[0,1]^3$, with dynamics: \vspace{-0.5em}
\begin{equation}\label{eq:lander}
\begin{aligned}
  \dot p_x&=v_x, & \quad \dot v_x&=-r_m u_0\sin\theta-r_s(u_1-u_2)\cos\theta,\\
  \dot p_y&=v_y, & \dot v_y&=r_m u_0\cos\theta-r_s(u_1-u_2)\sin\theta-g,\\
  \dot\theta&=\omega, & \dot\omega&=-r_I(u_1-u_2),
\end{aligned}
\vspace{-0.5em}
\end{equation}
where $g=0.2$ is known and $r:=(r_m,r_s,r_I) = (k_m/m, k_s/m, k_s/I)$ collects the unknown ratios of thrust to mass and
inertia. This system has the form \eqref{eq:greybox} with $d=12$ features,\vspace{-0.5em}
\begin{equation}\label{eq:lander-features}
  \phi(x,u)=\big[1,\ v_x,\ v_y,\ \omega,\ u_0\sin\theta,\ u_0\cos\theta,\ u_1\sin\theta,\
  u_1\cos\theta,\ u_2\sin\theta,\ u_2\cos\theta,\ u_1,\ u_2\big]^\top. \vspace{-0.5em}
\end{equation}
The lander must remain in the constraint region $|p_x|\!\le \!1.92$, $0\!\le\! p_y\!\le\! 3.42$,
$|v_x|,|v_y|\!\le\! 3.84$, $|\theta|\le\!\pi/2$, $|\omega|\le\! 1.92\pi$ for $T\!=\!4$~s. The dataset consists of five
trajectories of 20 samples each, with $\sigma\!=\!0.1$.

Each entry of $\Theta$ is either a known constant or a known multiple of one entry of $r$.
Therefore, $\tilde\Theta=\tilde\Theta_{\mathrm{c}}+Gr$ with known
$\tilde\Theta_{\mathrm{c}}\in\mathbb{R}^{72}$ and $G\in\mathbb{R}^{72\times 3}$. We place the Gaussian
prior $r\sim\mathcal{N}(\mu_r,\Sigma_r)$ with $\mu_r=(6,2,2)$, the simulated
value, and $\Sigma_r=0.5\,I_3$,
which gives $\tilde\Theta$ the mean $m_0=\tilde\Theta_{\mathrm{c}}+G\mu_r$ and the covariance
$G\Sigma_r G^\top$. This covariance has rank 3 and is not invertible. We therefore set
$\Sigma_0=G\Sigma_r G^\top+\gamma I_{72}$ with $\gamma=10^{-8}$, and using $dn=72$, we get $\beta_D=9.37$.

Grid-based methods become intractable as the state dimension grows. For this system, we therefore
use DeepReach \citep{bansal2021deepreach}, which represents the value function by a neural network
with sinusoidal activations. We train two networks: $\hat V_{\theta_1}$ with $\hat H$ and, as a
reference, $V_{\theta_2}$ with the true Hamiltonian $H$. The network $\hat V_{\theta_1}$ minimizes the loss
\begin{equation}\label{eq:deepreach}
\begin{aligned}
  \hat{\mathcal{L}}(\theta_1)={}&\sum_{j}\big|\hat V_{\theta_1}(x_j,0)-\ell(x_j)\big|\\
  &+\sum_{k}\Big|\min\Big\{\ell(x_k)-\hat V_{\theta_1}(x_k,t_k),\
  -\partial_t\hat V_{\theta_1}(x_k,t_k)+\hat H\big(x_k,\nabla_x\hat V_{\theta_1}(x_k,t_k)\big)\Big\}\Big|,
\end{aligned}
\vspace{-0.5em}
\end{equation}

\begin{figure}[t]
  \centering
  \includegraphics[width=\linewidth]{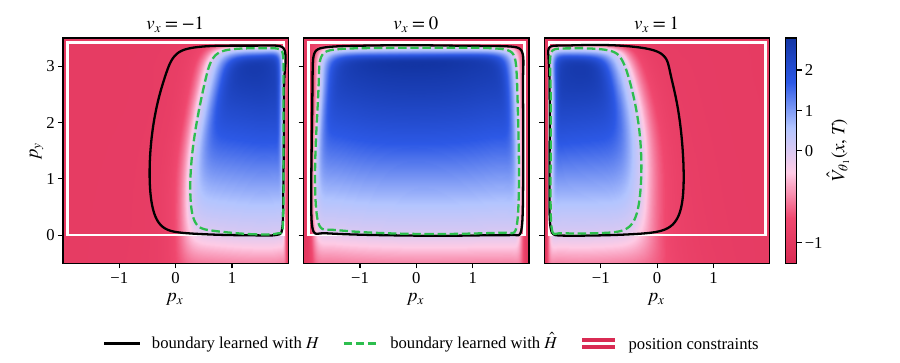}
  \vspace{-1.5em}
  \caption{Position slices of the learned safe sets of the lunar lander at $v_y, \theta, \omega=0$ for the horizon
  $T=4$~s under different values of $v_x$. The heat map is the value function learned with $\hat H$, on one color scale centered at zero.}
  \label{fig:lander-slices}
\vspace{-1em}
\end{figure}

\noindent and $V_{\theta_2}$ minimizes the loss $\mathcal{L}(\theta_2)$ obtained by replacing $\hat V_{\theta_1}$
and $\hat H$ in \eqref{eq:deepreach} with $V_{\theta_2}$ and $H$, respectively. The states $x_j$ and the
state--time pairs $(x_k,t_k)$ are sampled uniformly from the domain. The first sum penalizes
violations of the initial condition, and the second penalizes the residual of the HJ-VI. For
$\hat V_{\theta_1}$, we add a one-sided penalty on $\hat V_{\theta_1}(x, t)>\ell(x)$ at states sampled near the
position constraints; the exact solution incurs no such penalty, because $\hat V(x, t)\le\ell(x)$ for all $t \ge 0$. 

The
safety controllers of the two networks are
\begin{equation}\label{eq:controller}
  \hat u(x,t)\in\arg\max_{u\in U}\ \hat h\big(x,u,\nabla_x\hat V_{\theta_1}(x,t)\big),
\end{equation}
that is, the maximizer of \eqref{eq:pddh} at $p=\nabla_x\hat V_{\theta_1}(x,t)$, and
\begin{equation}\label{eq:controller-true}
  u^\star(x,t)\in\arg\max_{u\in U}\ \nabla_xV_{\theta_2}(x,t)^\top f(x,u).
\end{equation}

\begin{figure}[t]
  \centering
  \includegraphics[width=\linewidth]{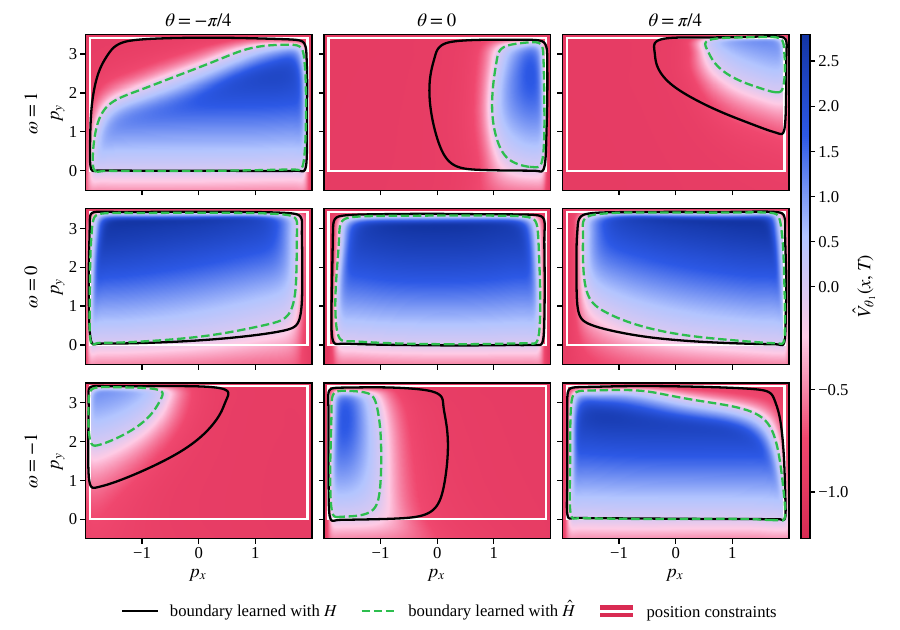}
  \vspace{-2em}
  \caption{Slices of the learned safe sets of the lunar lander at $v_x
\!=\!v_y\!=\!0$ for $T\!=\!4$~s, with $\theta$
  varying across the columns and $\omega$ across the rows. The heat map is the value function learned with
  $\hat H$, on one color scale centered at zero.}
  \label{fig:lander-slices-attitude}
\vspace{-1em}
\end{figure}

Figure~\ref{fig:lander-slices} shows three position slices of the learned safe sets under different horizontal velocities. The safe set learned with
$\hat H$ lies almost entirely inside the safe set learned with $H$. Figure~\ref{fig:lander-slices-attitude}
shows position slices under several values of $\theta$ and $\omega$. Both safe sets shrink when the lander rotates away
from the upright configuration, and the safe set learned with $\hat H$ again lies almost entirely inside the safe
set learned with $H$.

\begin{figure}[t]
  \centering
  \includegraphics[width=0.9\linewidth]{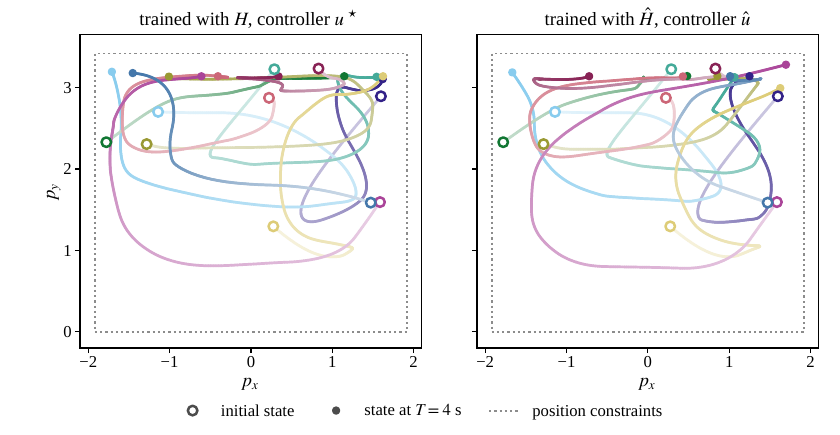}
  \vspace{-1em}
    \caption{Trajectories of the lunar lander over $T=4$~s from ten initial states in both learned safe
  sets, under $u^\star$ (left) and $\hat u$ (right). Each color corresponds to one initial state, and each
  line darkens as time increases. From the same initial states, the two controllers steer the lunar lander
  along different paths. While every trajectory stays inside the constraint region in both panels, the trajectories under the parametric DDH controller tend to be more conservative.}
  \label{fig:lander-traj}
\vspace{-1.5em}
\end{figure}

For the Monte Carlo validation, we draw initial states uniformly from the training domain and retain
the first 1000 that lie in each learned safe set. We then simulate the true dynamics for 4~s by forward Euler with a step of $0.01$~s under the
corresponding controller, with $t$ equal to the remaining horizon. Table~\ref{tab:mc} reports the
results, which describe one trained network for each Hamiltonian. With $\hat H$, 981 of the 1000
trajectories remain in the constraint region, against 946 with $H$. Figure~\ref{fig:lander-traj} shows
the trajectories from ten initial states that lie in both learned safe sets.

\begin{table}[h]
  \centering
  \small
  \caption{Monte Carlo validation of the lunar lander.}
  \vspace{0.5em}
  \label{tab:mc}
  \begin{tabular}{llc}
    \toprule
    Network trained with & Controller & Trajectories remaining in the constraint region\\
    \midrule
    $\hat H$ & $\hat u$ & 981 of 1000\\
    $H$ (reference) & $u^\star$ & 946 of 1000\\
    \bottomrule
  \end{tabular}
  \vspace{-1em}
\end{table}

\section{Conclusion \& Future Work}

\vspace{-0.5em}

We presented the parametric data-driven Hamiltonian for systems whose dynamics are linear in
unknown parameters through a known feature vector. From a Gaussian prior and noisy velocity
measurements, Bayesian linear regression gives a Gaussian posterior over the parameters. Taking the
worst case over a confidence ellipsoid of this posterior yields a lower bound on the Hamiltonian
that holds at all states and costates with probability at least $1-\delta$, without requiring a Lipschitz constant of the dynamics. The resulting safe set is contained in the true safe set with
the same probability. For control-affine features, the bound is evaluated as a second-order cone
program, so grid-based and neural PDE solvers can evaluate it at every queried state and costate efficiently.
On a two-state system, the data-driven safe sets were contained in the true safe set, whereas using
only the posterior mean produced safe sets that contain unsafe states. On a six-state lunar lander
with 100 data samples, 981 of 1000 simulated trajectories starting in the safe set learned with the
parametric data-driven Hamiltonian remained in the constraint region. The safe sets estimated by neural networks remain subject to approximation errors and therefore do not automatically inherit the theoretical containment guarantee. Future work should address these errors through calibration \citep{lin2024verification} or verification \citep{li2025certifiable}. Another direction is to use meta-learning to learn the feature vector and prior, both assumed known in this work, extending the applicability of parametric DDH to systems whose features are difficult to derive from first principles.

\vspace{-0.5em}
\subsection*{Generative AI Use}
\vspace{-0.5em}

Generative AI tools, including Claude (Anthropic) and ChatGPT (OpenAI), were used to assist with code development and debugging. These tools also helped the review of relevant literature to help reformulate the parametric data-driven Hamiltonian as a second-order cone program, as well as the drafting and polishing of the manuscript.

\acks{This work was supported by the Strategic Hub for International Research Collaboration project of Seoul National University (SNU).}

\bibliography{l4dc2027-sample}

\clearpage
\appendix

\section{Comparison Lemma for Value Functions}
\label{app:hj-comparison}

Consider the HJ-VI in \eqref{eq:hjvi} with initial condition $V(x,0)=\ell(x)$.

\begin{definition}[Viscosity subsolution and supersolution]
An upper semicontinuous function
$u:\mathbb{R}^n\times[0,T]\to\mathbb{R}$
is a \textit{viscosity subsolution} of \eqref{eq:hjvi} if for every
$\varphi\in C^1(\mathbb{R}^n\times (0, T))$
such that $u-\varphi$ has a local maximum of value zero at
$(x_0,t_0)\in\mathbb{R}^n\times(0, T)$,
\begin{equation}
\label{eq:viscosity-subsolution}
\min\bigl\{
\ell(x_0)-\varphi(x_0,t_0),\;
-\partial_t\varphi(x_0,t_0)
+H(x_0,\nabla_x\varphi(x_0,t_0))
\bigr\}\ge 0.
\end{equation}

A lower semicontinuous function
$u:\mathbb{R}^n\times[0,T]\to\mathbb{R}$
is a \textit{viscosity supersolution} of  \eqref{eq:hjvi} if for every
$\varphi\in C^1(\mathbb{R}^n\times(0, T))$
such that $u-\varphi$ has a local minimum of value zero at
$(x_0,t_0)\in\mathbb{R}^n\times(0, T)$,
\begin{equation}
\label{eq:viscosity-supersolution}
\min\bigl\{
\ell(x_0)-\varphi(x_0,t_0),\;
-\partial_t\varphi(x_0,t_0)
+H(x_0,\nabla_x\varphi(x_0,t_0))
\bigr\}\le 0.
\end{equation}

A continuous function is a \textit{viscosity solution} of \eqref{eq:hjvi} if it is both a viscosity subsolution
and a viscosity supersolution of the variational inequality.
\end{definition}

\begin{lemma}[Monotonicity with respect to the Hamiltonian]
\label{prop:hamiltonian-monotonicity}
Let $V_1,V_2$ be continuous viscosity solutions of
\eqref{eq:hjvi} with Hamiltonians $H_1$ and $H_2$,
respectively, with the same initial condition $V(x,0)=\ell(x)$. Assume that $H_1$ and $H_2$ satisfy 
\[
H_1(x,p)\le H_2(x,p)
\qquad
\forall \;(x,p)\in\mathbb{R}^n\times\mathbb{R}^n.
\]
Then
\[
V_1(x,t)\le V_2(x,t)
\qquad
\forall\; (x,t)\in\mathbb{R}^n\times[0,T].
\]
\end{lemma}

\begin{proof}
We show that $V_1$ is a viscosity subsolution of the HJ-VI associated with $H_2$.

Let $\varphi\in C^1(\mathbb{R}^n\times(0,T))$ be such that
$V_1-\varphi$ has a local maximum of value zero at
$(x_0,t_0)\in\mathbb{R}^n\times(0,T)$.
Since $V_1$ is a viscosity subsolution of the $H_1$ problem,
\[
\min\bigl\{
\ell(x_0)-\varphi(x_0,t_0),\;
-\partial_t\varphi(x_0,t_0)
+H_1(x_0,\nabla_x\varphi(x_0,t_0))
\bigr\}\ge 0.
\]
Using $H_1\le H_2$, we obtain
\begin{align*}
&\min\bigl\{
\ell(x_0)-\varphi(x_0,t_0),\;
-\partial_t\varphi(x_0,t_0)
+H_2(x_0,\nabla_x\varphi(x_0,t_0))
\bigr\}\\
&\quad\ge
\min\bigl\{
\ell(x_0)-\varphi(x_0,t_0),\;
-\partial_t\varphi(x_0,t_0)
+H_1(x_0,\nabla_x\varphi(x_0,t_0))
\bigr\}
\ge 0.
\end{align*}
Thus, $V_1$ is a viscosity subsolution of the $H_2$ equation, which satisfies $V_1(x,0)=\ell(x)=V_2(x,0)$. Note that by definition, $V_2$ is a viscosity supersolution of the $H_2$ equation. Since a viscosity supersolution is greater than or equal to a viscosity subsolution sharing the same initial condition, $V_1(x,t)\le V_2(x,t)$ for all
$(x,t)\in\mathbb{R}^n\times[0,T]$.
\end{proof}

\section{DeepReach Training Details}\label{app:deepreach}

Table~\ref{tab:train-shared} lists the settings shared by the two networks of
Section~\ref{sec:lander}, and Table~\ref{tab:train-diff} lists the settings in which they differ.

\begin{table}[htbp]
  \centering
  \caption{Settings shared by $\hat V_{\theta_1}$ and $V_{\theta_2}$.}
  \label{tab:train-shared}
  \begin{tabular}{lp{0.66\linewidth}}
    \toprule
    Setting & Value\\
    \midrule
    Network & sinusoidal activations, 3 hidden layers of 512 units\\
    Network input & $(t,p_x,v_x,p_y,v_y,\sin\theta,\cos\theta,\omega)$\\
    Sampling domain & $|p_x|\le 2$, $|v_x|,|v_y|\le 4$, $-0.5\le p_y\le 3.5$, $|\theta|\le\pi$,
      $|\omega|\le 2\pi$\\
    Horizon & $4$~s\\
    Optimizer & Adam, learning rate $2\times10^{-5}$, multiplied by $0.2$ at iteration $80{,}000$\\
    Stage 1 & fit of the initial condition with $90{,}000$ states at $t=0$ per iteration,
      $120{,}000$ iterations\\
    Stage 2 & initialized from stage 1, $120{,}000$ iterations, initial condition only for the
      first $1{,}000$\\
    Times in stage 2 & uniform, with an upper limit that grows linearly from $0$ to $4$~s between
      iterations $1{,}000$ and $60{,}000$\\
    \bottomrule
  \end{tabular}
\end{table}

\begin{table}[htbp]
  \centering
  \caption{Settings that differ between the two networks in stage 2. The first three rows count the
  training samples drawn anew at every iteration: the states $x_j$ at $t=0$ and the state--time pairs
  $(x_k,t_k)$ of \eqref{eq:deepreach}, sampled as in Table~\ref{tab:train-shared}, and the states of the
  one-sided penalty, drawn between the position limits and the edge of the sampling domain at times
  uniform over the horizon. For each penalty state, with probability $3/4$, $v_x$, $v_y$, and $\omega$
  are drawn from $[-0.6,0.6]$ and $\theta$ from $[-0.5,0.5]$. The HJ-VI residual is also evaluated at
  the states $x_j$. The true Hamiltonian $H$ is evaluated exactly: since $f$ is control-affine and $U$
  is a box, each control takes the bound selected by the sign of its coefficient.}
  \label{tab:train-diff}
  \begin{tabular}{p{0.42\linewidth}p{0.24\linewidth}p{0.2\linewidth}}
    \toprule
    & $\hat V_{\theta_1}$ (with $\hat H$) & $V_{\theta_2}$ (with $H$)\\
    \midrule
    Initial-condition states per iteration & $140{,}000$ & $35{,}000$\\
    HJ-VI state--time samples per iteration & $400{,}000$ & $100{,}000$\\
    One-sided penalty states per iteration & $8{,}192$, weight $1$ & none\\
    Hamiltonian evaluation & $40$ Adam iterations, projected onto $U$ & exact\\
    \bottomrule
  \end{tabular}
\end{table}

\section{Lunar Lander Model}\label{app:lander}

In the lunar lander of Section~\ref{sec:lander}, the main thruster produces the force $k_m u_0$ along the body's vertical axis, and the side thrusters produce the forces $k_s u_1$ and $k_s u_2$ in opposite directions along its horizontal axis, with a moment arm of $1$. Table~\ref{tab:lander-params} lists the physical parameters.

Newton's laws for these forces and gravity give \eqref{eq:lander}, which is linear in $r=(k_m/m,\,k_s/m,\,k_s/I)$ but not in $m$ and $I$. With the features \eqref{eq:lander-features}, the dynamics take the form \eqref{eq:greybox}:
\setcounter{MaxMatrixCols}{12}
\begin{equation}\label{eq:lander-theta}
  \begin{bmatrix}\dot p_x\\ \dot v_x\\ \dot p_y\\ \dot v_y\\ \dot\theta\\ \dot\omega\end{bmatrix}
  =\underbrace{\begin{bmatrix}
    0&1&0&0&0&0&0&0&0&0&0&0\\
    0&0&0&0&-r_m&0&0&-r_s&0&r_s&0&0\\
    0&0&1&0&0&0&0&0&0&0&0&0\\
    -g&0&0&0&0&r_m&-r_s&0&r_s&0&0&0\\
    0&0&0&1&0&0&0&0&0&0&0&0\\
    0&0&0&0&0&0&0&0&0&0&-r_I&r_I
  \end{bmatrix}}_{\Theta^\top}
  \underbrace{\begin{bmatrix}1\\ v_x\\ v_y\\ \omega\\ u_0\sin\theta\\ u_0\cos\theta\\ u_1\sin\theta\\
    u_1\cos\theta\\ u_2\sin\theta\\ u_2\cos\theta\\ u_1\\ u_2\end{bmatrix}}_{\phi(x,u)}.
\end{equation}
The entries $1$ and $-g$ are known constants and form $\tilde\Theta_{\mathrm{c}}$. Each remaining nonzero
entry is $\pm r_m$, $\pm r_s$, or $\pm r_I$, and these entries give the term $Gr$ in
$\tilde\Theta=\tilde\Theta_{\mathrm{c}}+Gr$ of Section~\ref{sec:lander}.

\begin{figure}[t]
  \centering
  \includegraphics[width=0.3\linewidth]{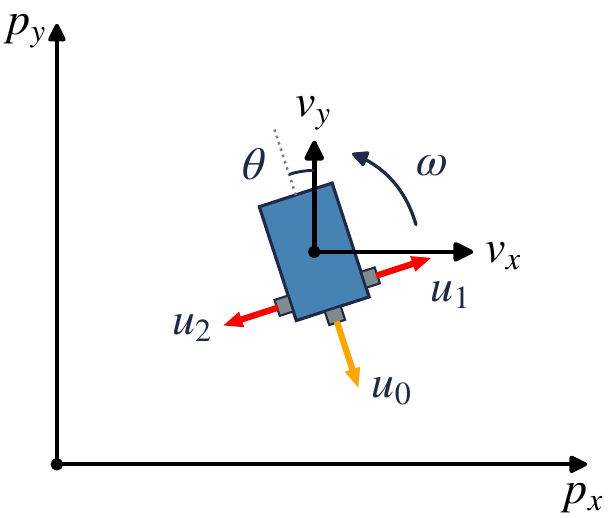}
  \caption{States and thruster inputs of the lunar lander.}
  \label{fig:lander-sketch}
\end{figure}

\begin{table}[!ht]
  \centering
  \caption{Physical parameters of the lunar lander. Only $g$ is known to the method; $m$, $I$, $k_m$,
  and $k_s$ enter the dynamics only through the ratios $r$, whose simulated value is $(6,2,2)$.}
  \label{tab:lander-params}
  \begin{tabular}{llc}
    \toprule
    Parameter & Meaning & Simulated value\\
    \midrule
    $m$ & mass & $1$\\
    $I$ & moment of inertia about the center of mass & $1$\\
    $k_m$ & thrust parameter of the main thruster & $6$\\
    $k_s$ & thrust parameter of each side thruster & $2$\\
    $g$ & gravitational acceleration & $0.2$\\
    \bottomrule
  \end{tabular}
\end{table}

\end{document}